\documentclass[10pt,conference]{IEEEtran}
\usepackage{listings}
\IEEEoverridecommandlockouts

\usepackage[numbers,sort&compress]{natbib}
\usepackage{hyperref}
\hypersetup{
    colorlinks=true,    
    citecolor=black,    
    linkcolor=black,    
    urlcolor=SkyBlue,      
    pdfborder={0 0 0}   
}
\usepackage{amsmath,amssymb,amsfonts,array,tabularx}
\usepackage{xurl}
\usepackage{algorithmic}
\usepackage{graphicx}
\usepackage{textcomp}
\usepackage{booktabs}
\usepackage[dvipsnames]{xcolor}
\usepackage{subcaption}
\usepackage{multirow}
\begin{document}

\title{LLM-Enhanced Hierarchical Heterogeneous Graph Representation Learning for Malicious Python Package Detection
}

\author{
\IEEEauthorblockN{Hang Gao$^{1,2,\ast}$,
Xiaoyu Chen$^{1,2,\ast}$,
Baoquan Cui$^{1,2}$,
Zhen Tang$^{1,2}$,
Peng Qiao$^{1,2}$,
Fengge Wu$^{1,2,\dagger}$,
Jian Zhang$^{1,2}$}
\IEEEauthorblockA{$^1$Institute of Software, Chinese Academy of Sciences}
\IEEEauthorblockA{$^2$University of Chinese Academy of Sciences}
\IEEEauthorblockA{\{gaohang, chenxiaoyu2025, qiaopeng, fengge\}@iscas.ac.cn;
\{cuibq, zj\}@ios.ac.cn; tangzhen12@otcaix.iscas.ac.cn \\
$^\ast$Equal contribution \quad
$^\dagger$Corresponding author: \href{mailto:fengge@iscas.ac.cn}{fengge@iscas.ac.cn}}
}

\maketitle

\begin{abstract}
Malicious Python packages have become a major threat to modern software supply chain ecosystems due to the widespread adoption of open-source repositories such as PyPI. Existing learning-based detection methods struggle to capture the hierarchical organization and heterogeneous interactions among different program entities. Although Large Language Models (LLMs) have demonstrated remarkable capabilities in code understanding and semantic reasoning, they are rarely integrated with structural program representations for fine-grained malicious behavior analysis. In this paper, we propose an LLM-enhanced hierarchical heterogeneous graph representation learning framework for malicious Python package detection. The proposed framework constructs a hierarchical heterogeneous code graph that explicitly models heterogeneous code entities, together with different types of structural dependencies. To further enrich code representations, LLMs are leveraged to infer function-level semantic roles, introducing an additional layer of semantic heterogeneity. Based on this graph, we develop a hierarchical heterogeneous graph neural network that performs type-aware message passing over different node and edge categories, enabling effective modeling of malicious behavior propagation and accurate package-level classification. Furthermore, the proposed framework incorporates a function-level attribution mechanism which, combined with LLM reasoning, automatically identifies suspicious functions and localizes fine-grained malicious behaviors without requiring human expert intervention. Extensive experiments on real-world datasets demonstrate that the proposed framework consistently outperforms traditional machine learning methods, graph-based detectors, and state-of-the-art LLMs across packages with varying sizes and dependency complexities, while providing accurate, robust, and interpretable malicious behavior localization. The replication package is available at: \url{https://github.com/xxy33/malware}
\end{abstract}


\section{Introduction}

With the continued shift of modern software development toward modularity and code reuse, Open Source Software (OSS) has become a fundamental pillar of digital infrastructure. As one of the most active software ecosystems, the Python Package Index (PyPI) hosts more than 500,000 packages and serves billions of downloads each month~\cite{pypi_stats}. However, the openness and highly interconnected nature of the ecosystem have also made it a prime target for software supply chain attacks. High-profile security incidents, such as the SolarWinds compromise~\cite{solarwinds} and the Log4j vulnerability~\cite{log4j}, have demonstrated that the compromise of a single component can trigger cascading effects throughout the entire software supply chain. Although industry initiatives such as SLSA~\cite{slsa} and in-toto~\cite{torres2019intoto} have been proposed to improve supply chain security practices, attacks targeting package management ecosystems remain prevalent. Adversaries exploit the trust model of software repositories through techniques such as typosquatting, dependency confusion, and malicious installation script injection~\cite{taylor2020typosquatting,neupane2023beyond} to steal sensitive credentials, establish reverse shells, and execute arbitrary malicious code~\cite{zimmermann2019npm,vu2023badsnakes}.

To mitigate these threats, a variety of rule-based detection approaches have been proposed~\cite{ladisa2023sok}. These methods typically rely on predefined signatures, suspicious API usage patterns, heuristic rules, or manually crafted indicators of compromise (IoCs) to identify malicious packages~\cite{sejfia2022practical,duan2021towards}. While effective against known attack patterns, they often struggle to cope with the rapidly evolving landscape of software supply chain attacks. Malicious packages frequently employ code obfuscation, dynamic code loading, environment-aware execution, and indirect invocation mechanisms to evade static rule-based detection~\cite{ohm2020backstabber,vaidya2019security}. Furthermore, the highly dynamic nature of Python, including features such as \texttt{eval}, \texttt{exec}, reflection, and runtime dependency resolution, further complicates the construction of precise detection rules. Consequently, rule-based approaches generally suffer from limited generalization capability and tend to exhibit high false-positive and false-negative rates when confronted with previously unseen attacks or carefully disguised malicious behaviors.

In recent years, learning-based approaches have attracted considerable attention for malicious package detection due to their ability to automatically learn discriminative malicious behavior patterns from large-scale software package corpora~\cite{guo2024empirical}. Existing studies have explored a variety of techniques, including traditional machine learning classifiers~\cite{ohm2022feasibility}, Graph Neural Networks (GNNs)~\cite{li2026pvdetector,zhou2019devign}, and LLMs~\cite{wyss2025evaluating}. Among them, GNNs are capable of capturing structural dependencies within programs, while LLMs demonstrate strong capabilities in code semantic understanding and natural language reasoning. Consequently, integrating GNNs and LLMs for code security analysis seems to be a promising research direction. In fact, relevant cutting-edge explorations have already begun~\cite{gao2025malguard,linredetect}.


Such explorations open up a largely unexplored yet promising research direction for malicious Python package detection. However, former works that integrate GNNs with LLMs still rely on homogeneous graph representations, where different types of program entities and relations are modeled uniformly. This limitation is particularly pronounced in Python package ecosystems, where malicious behaviors often emerge through interactions among heterogeneous components, including packages, files, functions, installation scripts, imported modules, and external dependencies. Attack logic is rarely confined to a single code fragment; instead, it propagates across multiple levels of program organization through import relations, function invocations, installation-time execution, and dynamic code loading. Modeling these interactions using homogeneous graphs makes it difficult to accurately characterize malicious behavior propagation. 

\begin{figure}[h]
    \centering
    \begin{subfigure}[b]{0.15\textwidth}
        \centering
        \includegraphics[width=\textwidth]{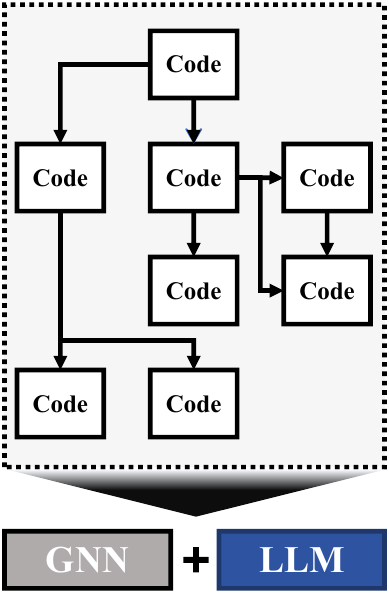}
        \caption{}
        \label{fig:paradigm_a}
    \end{subfigure}
    \hfill
    \begin{subfigure}[b]{0.15\textwidth}
        \centering
        \includegraphics[width=\textwidth]{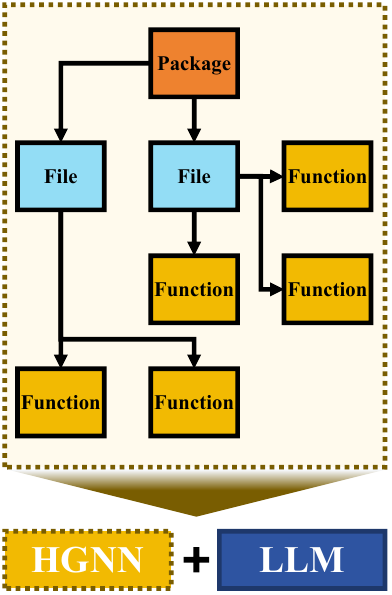}
        \caption{}
        \label{fig:paradigm_b}
    \end{subfigure}
    \hfill
    \begin{subfigure}[b]{0.16\textwidth}
        \centering
        \includegraphics[width=\textwidth]{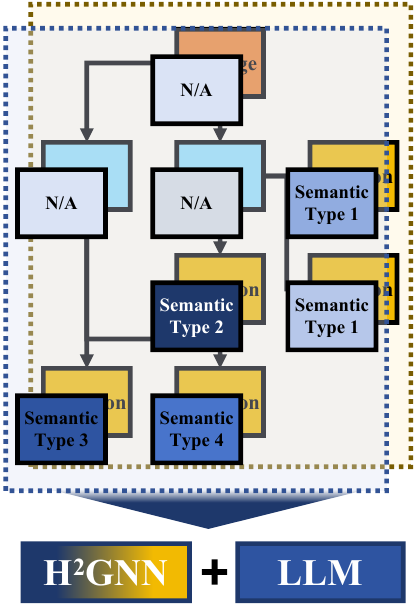}
        \caption{}
        \label{fig:paradigm_c}
    \end{subfigure}
    \caption{Three types of code security analysis paradigms integrating GNNs and LLMs. (a) Existing approaches directly process code graphs, failing to capture heterogeneous characteristics. (b) Introducing heterogeneous graph representation learning enables the modeling of explicit heterogeneity, yet latent heterogeneity remains largely unexplored. (c) Our paradigm constructs a hierarchical heterogeneous graph and corresponding Hierarchical Heterogeneous GNN (H$^2$GNN) to effectively capture both explicit and latent heterogeneity.}
    \label{fig:ins}
\end{figure}

A natural solution is to incorporate heterogeneous graph learning into GNN-LLM frameworks for malicious package detection. To the best of our knowledge, such paradigm has not been explored for malware detection yet. However, existing heterogeneous graph methods rely on explicitly defined node and edge types, whereas the heterogeneity in malware is partially latent and behavior-driven. As a result, they cannot effectively capture the implicit semantics of malicious behaviors. Figure \ref{fig:paradigm_a} and \ref{fig:paradigm_b} provides an intuitive illustration.

To address these challenges, we propose a \textit{\textbf{H}ierarchical \textbf{H}eterogeneous \textbf{G}raph representation learning framework enhanced by \textbf{L}LMs for \textbf{M}alicious package detection and attack-chain localization} (\textbf{H$^2$GLM}). Specifically, H$^2$GLM is designed for Python package ecosystems and constructs a hierarchical heterogeneous code graph that explicitly models different types of code entities, including packages, files, and functions, as well as heterogeneous dependencies such as containment, import, and function-call relations. In addition, LLMs are leveraged to infer function-level semantic types, introducing latent semantic heterogeneity into the graph. Based on this graph, we develop a hierarchical heterogeneous graph neural network that captures malicious behavior propagation and attack logic across the code structure. Furthermore, by combining graph attribution analysis with LLM-based reasoning, H$^2$GLM can fully automatically localize malicious functions, identify critical attack paths, enabling accurate, efficient, and interpretable malicious package analysis.

The main   contributions are listed as follows:

\begin{itemize}

\item We propose H$^2$GLM, a learning-based framework for malicious package detection and attack-chain localization in Python package ecosystems. By integrating the strengths of GNNs and LLMs while jointly modeling the explicit and implicit heterogeneity of malicious code, H$^2$GLM automatically identifies malicious functions and localizes malicious behaviors.

\item We develop a novel hierarchical heterogeneous graph neural network tailored for code analysis, which jointly models different program entities, heterogeneous dependencies, and LLM-derived semantics.

\item Extensive experiments on benchmark datasets and real-world PyPI packages demonstrate the effectiveness of H$^2$GLM in malicious package detection and attack-chain localization, while uncovering multiple real-world malicious packages acknowledged by PyPI maintainers.

\end{itemize}

\begin{figure*}[h!]
    \centering
    \includegraphics[width=1\textwidth]{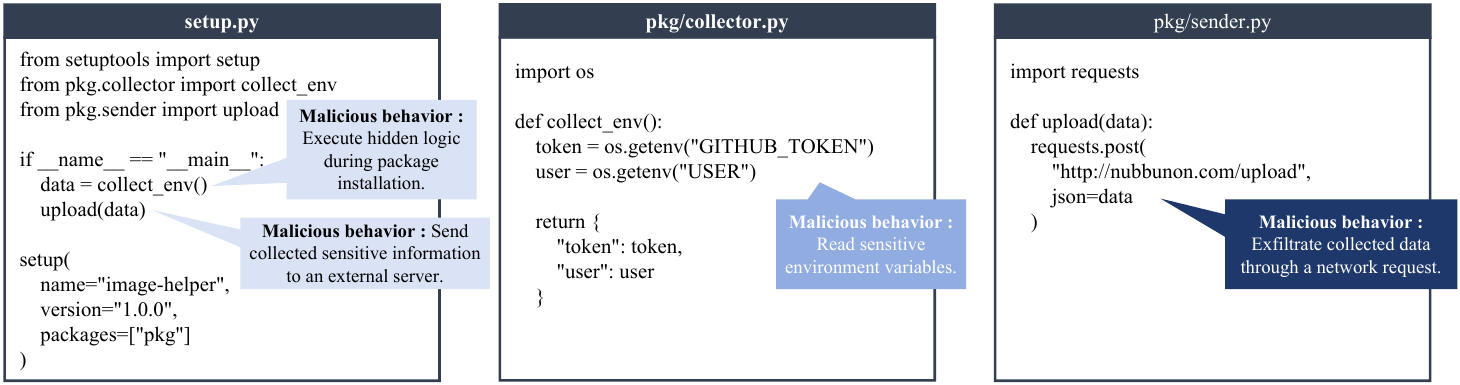}
    
    \caption{The figure shows a malicious Python package where attacks span various stages.
    }
    \vskip -0.1in
    
    \label{fig:mexample}
\end{figure*}

\section{Background}

Malware detection aims to identify hidden malicious logic embedded in packages, scripts, or modules in order to mitigate software supply chain attacks~\cite{ohm2020backstabber}. Due to the openness of Python package ecosystems such as PyPI and the dynamic nature of the Python language, attackers commonly exploit package distribution channels through dependency confusion, typosquatting, malicious installation script injection, and dynamic code loading~\cite{neupane2023beyond,zimmermann2019npm}. Attacks against PyPI have become increasingly severe, as malicious packages can be rapidly distributed to a large number of downstream users and projects once published~\cite{vu2023badsnakes,guo2024empirical}. These malicious behaviors may involve credential theft, remote code execution, backdoor implantation, environment inspection, and unauthorized network communication~\cite{duan2021towards,ferreira2021containing}. In recent years, learning-based approaches have been increasingly adopted for Python malware detection, as they can automatically learn malicious behavioral patterns from large-scale Python code repositories and generally exhibit stronger generalization ability than manually crafted rules and signature-based methods~\cite{sejfia2022practical,sun2024plus}.

Python packages inherently contain rich structural and semantic information; consequently, GNNs and LLMs have become widely adopted in learning-based code security analysis~\cite{feng2020codebert,guo2021graphcodebert}. Graph-based representations can explicitly model relationships among program entities~\cite{zhou2019devign,cheng2021deepwukong}. Meanwhile, LLMs possess strong capabilities in Python code comprehension, semantic reasoning, and behavioral interpretation, enabling them to analyze functional roles, infer execution intent, and generate natural-language descriptions of suspicious behaviors~\cite{zhang2025cerebro}. However, malicious behaviors in Python packages often rely on heterogeneous program structures and diverse dependency relationships~\cite{ladisa2023sok}. As a result, relying solely on local code fragments or homogeneous graph representations is often insufficient to fully characterize malware behaviors~\cite{zhang2025cerebro,huang2024spiderscan}. Therefore, effectively providing models with richer, more structured, and semantically meaningful program context remains a fundamental challenge in learning-based malware analysis, particularly for GNN-LLM-based methods.

\section{Motivation}

\subsection{Motivating Example}

Consider the malicious Python package illustrated in Fig. \ref{fig:mexample}. The attack is initially triggered by an installation script, which subsequently imports several utility modules and invokes a function responsible for collecting environment information. The collected data is then forwarded to another function that establishes an external network connection, ultimately resulting in credential theft and data exfiltration.

From a security perspective, the attack objective is relatively clear. However, accurately identifying such an attack chain remains challenging for learning-based analysis systems. The primary reason is that the malicious logic is not concentrated within a single code fragment. Instead, it is distributed across multiple program entities, including packages, files, and functions, which are organized at different levels of abstraction. These entities are further connected through various dependency relationships, such as containment, import, and function invocation. Moreover, different functions often play distinct semantic roles, including environment inspection, credential access, and network communication.

Ideally, a malicious code analysis system should jointly leverage these heterogeneous sources of information to understand attack behaviors. However, existing LLM-enhanced GNN approaches still pay limited attention to the heterogeneous nature of program representations. As a result, they often fail to fully integrate semantic and structural information, making it difficult to accurately reconstruct complete attack behaviors and identify the truly malicious components.

\subsection{Observations}

\textbf{ \textit{1. Malicious Behaviors are Inherently Heterogeneous:}}
Malicious behaviors rarely reside within a single function or file. Instead, they emerge through interactions among heterogeneous program entities. Attack logic often spans multiple levels of software organization and propagates through diverse dependency relationships. Therefore, accurately understanding malicious behaviors requires jointly modeling program hierarchies, dependency structures, and functional semantics.

\textbf{ \textit{2. Semantic Heterogeneity is Not Directly Extractable:}}
In malware detection, security-critical heterogeneity is often implicit in semantic roles rather than explicit program structures. Functions for environment inspection, credential access, dynamic execution, and network communication may appear syntactically similar but play different roles in malicious behavior propagation. Such implicit heterogeneity cannot be directly handled by conventional HGNNs, which require predefined node or edge types before processing.

\textbf{ \textit{3. LLMs Lack Explicit Heterogeneous Context:}}
LLMs have recently demonstrated remarkable capabilities in code understanding and semantic reasoning. However, their attention mechanisms and sequential processing paradigm inherently limit their ability to perceive large-scale and highly interconnected program structures. Consequently, when a rich heterogeneous context is unavailable, LLMs must infer attack logic from incomplete information, which may lead to inaccurate behavior understanding and unreliable attack-chain reconstruction.

\subsection{Key Ideas}

Based on the above observations, we derive the following key ideas when designing H$^2$GLM.

\textbf{\textit{1. Preserve Heterogeneity as Much as Possible:}}
Rather than simplifying software packages into homogeneous representations, we explicitly preserve heterogeneous characteristics at multiple levels, including type-level heterogeneity, functionality-level heterogeneity, and heterogeneous dependency relations. By maintaining these diverse sources of information, we provide a richer contextual foundation for downstream malicious behavior analysis.



\textbf{\textit{2. Automatically Discover Heterogeneity:}}
Since functional heterogeneity is not directly observable from program structures, it must be explicitly inferred from code semantics. To this end, we leverage LLMs to analyze the source code of each function and actively discover its functional role, such as environment inspection, file-system operation, credential access, dynamic execution, or network communication. The inferred roles and generated semantic descriptions are then introduced as an additional layer of heterogeneity, enabling the model to capture security-critical functional differences beyond explicit structural types.

\textbf{\textit{3. Use Heterogeneous Features to Guide LLM Reasoning:}}
Existing studies typically employ LLMs as semantic encoders or post-hoc explanation tools. In contrast, we utilize the preserved heterogeneous structural and semantic information as contextual guidance for LLM reasoning. By jointly exploiting structural heterogeneity and semantic heterogeneity, LLMs can more accurately understand malicious behavior propagation, identify critical attack paths, and perform fine-grained malicious function localization.

\section{Method}
\begin{figure*}[h!]
    \centering
    \includegraphics[width=1\textwidth]{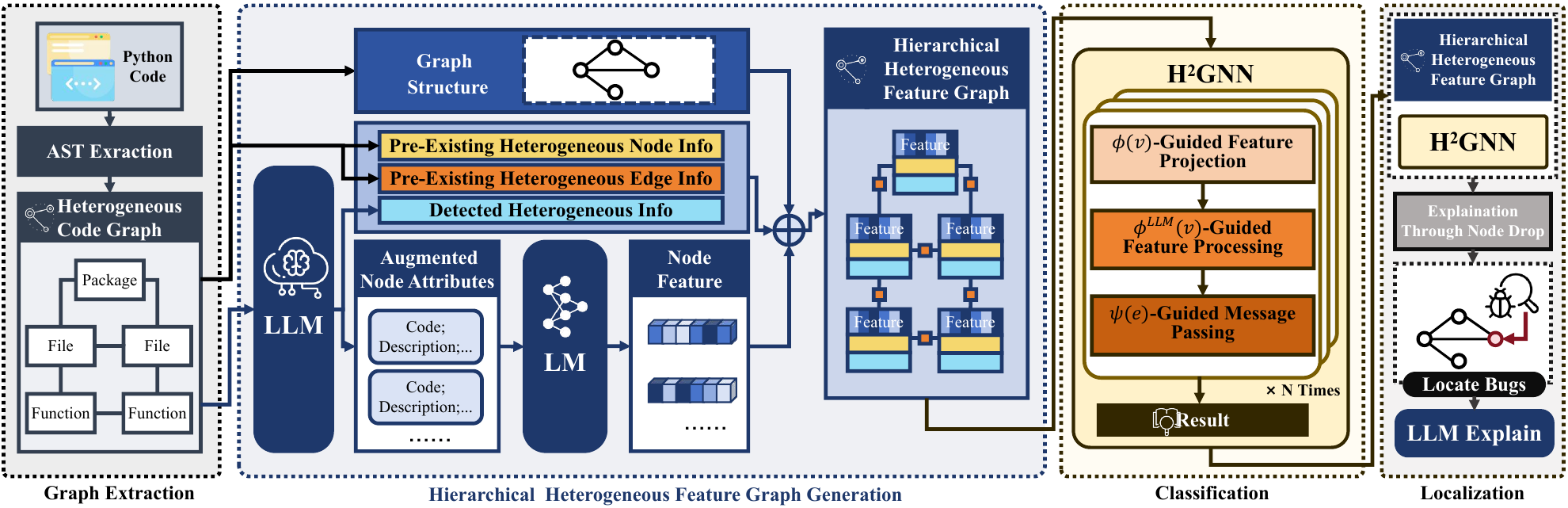}
    
    \caption{The framework of H$^2$GLM, consisting of four components:1) Graph Extraction, which constructs code data into a heterogeneous graph; 2) Hierarchical Heterogeneous Feature Graph Generation, which leverages LLMs to transform code data into a hierarchical heterogeneous graph to fully characterize the features contained within the code; 3) Classification, which utilizes a heterogeneous GNN specifically designed for malicious code detection to classify code and pinpoint issues; and 4) Localization, achieved through explaining the utilized neural network modules.
    }

    \vskip -0.1in
    
    \label{fig:example}
\end{figure*}


The proposed methodology consists of four components: Graph Extraction, Hierarchical Heterogeneous Feature Graph Generation, Classification, and Localization, as illustrated in Fig.~\ref{fig:example}. The following sections provide a detailed description of each component.

\subsection{Graph Extraction}
This part extracts the heterogeneous code graph $\mathcal{G} = (\mathcal{V}, \mathcal{E}, \phi, \psi)$, where $\mathcal{V}$ is the set of nodes. Each node $v$ in $\mathcal{V}$ corresponds to an extracted node from the ASTs of the package, including three types: package, file, and function. The ASTs are generated utilizing Python's built-in module.  $\phi$ is a mapping function that outputs each node's type. For each node $v$, we extract its description as its attribute. If the node is a Function, the description consists of the original source code. For File nodes, an LLM creates a short summary by looking at the descriptions of all functions in that file and the main project files, such as the root documentation, the project manifest, and the root module's docstring. For Package nodes, the LLM creates a summary only after all the file descriptions are ready. It uses those file descriptions along with the main project files to explain the package's purpose; if the main project files are missing, it simply uses the file descriptions to infer the summary instead. 

Each edge $e$ in $\mathcal{E}$ represents a relationship between the nodes, including three types: contain, call, and import. $\psi$ is an edge-type mapping function. Call edges are extracted directly from the ASTs. As for import edges, we first scan the package directory to create a lookup table for resolving imports. During the subsequent syntax-tree traversal, we identify import edges by matching import statements against this table. Simultaneously, we establish contain edges by linking each package to its files and each file to its defined functions.

\subsection{Hierarchical Heterogeneous Feature Graph Generation}

Based on the acquired $\mathcal{G}$, this part constructs a Hierarchical Heterogeneous Feature Graph, denoted as $\mathcal{G}^{\text{H}} = (\mathcal{V}, \mathcal{E}, \phi, \phi^{\text{LLM}}, \psi)$. Compared to the existing heterogeneous information (node types $\phi$ and edge types $\psi$) in the original graph $\mathcal{G}$, $\mathcal{G}^{\text{H}}$ introduces a novel heterogeneous information layer $\phi^{\text{LLM}}$. Together with $\phi(v)$, it provides the essential node-type data required for the subsequent construction of the hierarchical heterogeneous code graph. Specifically, $\phi^{\text{LLM}}$ utilizes an LLM to further process the content of each code function to dynamically detect its specific functional node type and generate a semantic description of its functionality. This process is formally defined as follows:
\begin{equation}
    (\phi^{\text{LLM}}(v) , d_{v}) = \text{LLM}(c_{v}),
\end{equation}
$c_{v}$ denotes the code context associated with node $v$. For File and Package nodes, $c_{v}=\text{NULL}$, since they do not directly correspond to executable code. $\phi^{\text{LLM}}(v)$ denotes the functional type assigned by the LLM, which is only applicable to Function nodes. The semantic description $d_{v}$ represents the functionality of the corresponding code for Function nodes, whereas for File and Package nodes, $d_{v}$ denotes the hierarchical summaries generated during the Graph Extraction step.
The LLM takes the source code as input and dynamically derives its specific category through analysis, rather than relying on pre-defined labels. Furthermore, $\phi^{\text{LLM}}(v)$ is exclusively applied to "Function" nodes, while "Package" and "File" nodes are excluded from this classification and default to the "N/A" category. In our practical implementation, we utilized the LLM to conduct a centralized scan across 100 packages in the target dataset to automatically generate and output a specific set of function types. Due to space constraints, we provide a detailed description of the prompts and their associated contents in the released implementation (please refer to the replication package link in the Abstract).

After obtaining the rich heterogeneous information mentioned above, to facilitate downstream GNN processing, we encode the relevant information of each node in $\mathcal{G}^{\text{H}}$ into a feature vector. This procedure can be defined as:
\begin{equation}
    h_v = \text{LM}^{\text{STF}}(d_{v}) \, || \, \text{LM}^{\text{CB}}(c_{v}),
\end{equation}
$h_v$ denotes the extracted feature vector for node $v$, and $||$ represents the feature concatenation operation. $\text{LM}^{\text{STF}}(\cdot)$ refers to a pre-trained Sentence Transformer model, which is responsible for transforming the generated textual description $d_{v}$ into a dense vector representation; $\text{LM}^{\text{CB}}(\cdot)$ denotes a CodeBERT model, which encodes the source code snippet $c_{v}$ into a semantically meaningful feature vector. During the feature extraction stage, we select these task-specific language models over general-purpose LLMs because our objective requires specialized feature representations optimized for structural and semantic code understanding. The extensive generalized knowledge and generative capabilities inherent in general LLMs introduce unnecessary computational overhead in this specific task without yielding commensurate performance benefits.

\subsection{Classification}

To facilitate rapid and efficient analysis on $\mathcal{G}^{\text{H}}$, we construct a novel \textit{\textbf{H}ierarchical \textbf{H}eterogeneous \textbf{G}raph \textbf{N}eural \textbf{N}etwork} (H$^{2}$GNN) tailored for analyzing the highly heterogeneous information within codes. Each layer of this specialized H$^{2}$GNN comprises three components: $\phi(v)$-guided feature projection, $\phi^{\text{LLM}}(v)$-guided feature processing, and $\psi(e)$-guided message passing. These components are detailed below.

\paragraph{$\phi(v)$-Guided Feature Projection}

The three node types in $\mathcal{G}^{\text{H}}$ possess feature vectors of different dimensionalities due to their distinct contents; therefore, we apply a type-specific feature alignment transformation. Specifically, for each node $v \in \mathcal{V}$, its original feature vector $h_v$ is projected into a unified latent space through a learnable type-dependent projection matrix:
\begin{equation}
h_v'^{(l)} = W^{(l)}_{\phi(v)} h_v , (l=1)
\label{eq:1}
\end{equation}
where $\phi(v)$ denotes the type of node $v$, $W^{(l)}_{\phi(v)} \in \mathbb{R}^{F_{\phi(v)} \times F'}$ is the learnable projection matrix associated with node type $\phi(v)$, and $l$ denotes the layer index. Equation~\ref{eq:1} corresponds to the projection operation in the first layer.
This operation maps the heterogeneous feature representations of packages, files, and functions into a common latent space of dimension $F'$, thereby enabling subsequent cross-type message passing and representation learning.

For the subsequent H$^{2}$GNN layers, since the node representations have already been projected into a unified feature space, the $\phi(v)$-guided message passing operates on projection matrices of identical dimensionality. Specifically, for $l$-th layer that $l > 1$, the projection matrix is defined as
$
W^{(l)}_{\phi(v)} \in \mathbb{R}^{F' \times F'}
$.

\paragraph{$\phi^{\text{LLM}}(v)$-Guided Feature Processing}
 
Similar to the $\phi$-guided feature projection, the computation process of the $\phi^{\text{LLM}}$-guided feature processing can be formally expressed as follows:
\begin{equation}
h_v''^{(l)} = W_{\phi^{\text{LLM}}(v)}^{(l)} h_v'^{(l)} ,
\end{equation}
where $W_{\phi^{\text{LLM}}(v)}^{(l)}\in \mathbb{R}^{F' \times F'}$ denotes a learnable matrix associated with the content type $\phi^{\text{LLM}}(v)$. This content-aware processing enables the model to learn specialized representations according to different node functional types.

\paragraph{$\psi(e)$-Guided Message Passing.}

This part performs message passing guided by the edge type $\psi(e)$ to propagate information across the code graph structure, ensuring each edge type governs distinct information flow.

For a target node $v$, the message aggregation is computed as:
\begin{equation}
    m_{v}^{(l)} = \frac{1}{|\mathcal{N}(v)|}\sum_{u \in \mathcal{N}(v)} \alpha_{\psi(e_{uv})}^{(l)} \cdot W_{\psi(e_{uv})}^{(l)} \cdot h''^{(l)}_{u},
\end{equation}
where $W_{\psi(e_{uv})}$ is a learnable weight matrix specific to the edge type $\psi(e_{uv})$, and $\alpha_{\psi(e_{uv})}$ is a learnable relation-specific coefficient that emphasizes critical relation types indicative of malicious behaviors. The node embedding is then updated as:
\begin{equation}
    h^{(l+1)}_{v} = \sigma \left( m_{v}^{(l)} + \Theta^{(l)} \cdot h''^{(l)}_{v} \right),
\end{equation}
where $\Theta$ is the weight matrix for the residual connection, and $\sigma$ denotes a non-linear activation function. We stack $L=2$ layers of this heterogeneous message passing to enable multi-hop information propagation across the hierarchical structure.

After $L$ layers of propagation, we obtain the final node representations $H^{(L)}$. To perform the graph-level classification for the Python package, we apply a global mean pooling operation $\text{READOUT}(\cdot)$ over the node representations to generate a graph embedding $h_{\mathcal{G}}$. This vector is subsequently fed into a Multi-Layer Perceptron $\text{MLP}(\cdot)$ to predict the classification $\hat{y}$ of maliciousness:
\begin{equation}
    \hat{y} = \text{MLP} \left( \text{READOUT}(\{h^{(L)}_{v} \mid v \in \mathcal{V}\}) \right).
\end{equation}

\subsection{Localization}
\label{subsec:Localization}
Beyond package-level malware detection, our framework further supports fine-grained localization of malicious behaviors. To achieve such a goal, we first select the packages that are classified as malicious and identify the functions that are highly associated with malicious behaviors. 

Let $v^f$ denote a node whose type is a function in $\mathcal{G}^{\text{H}}$.
For each $v^f$ in $\mathcal{G}^{\text{H}}$, we remove it from the graph one at a time and observe the resulting change in the model's predicted malicious probability. The underlying intuition is that, if a function is closely associated with malicious behavior, removing it from the $\mathcal{G}^{\text{H}}$ should weaken the malicious semantic signals it contributes, thereby reducing the model's confidence in classifying the package as malicious. By measuring the variation in the predicted malicious probability after the removal of each function node, we can quantify its contribution to the malicious prediction and identify the function nodes that are most strongly associated with malicious behavior, thereby conduct the explanation of the H$^2$GNN. We formally measure each $v^f$'s association with malicious behavior as $s(v^f)$.

$s(v^f)$ can be represented as follows:
\begin{equation}
s(v^f) = \mathcal{M}\big(\text{H}^{2}\text{GNN}(\mathcal{G}^{\text{H}})\big) - \mathcal{M}\big(\text{H}^{2}\text{GNN}(\mathcal{G}^{\text{H}}\setminus v^f)\big).
\end{equation}
$\text{H}^{2}\text{GNN}(\cdot)$ denotes our proposed GNN. $\mathcal{M}(\cdot)$ returns the probability assigned to the malicious class. $\mathcal{G}^{\text{H}}\setminus v^f$ denotes $\mathcal{G}^{\text{H}}$ removing $v^f$ together with all its incident edges.

Next, we rank all function nodes according to their malicious association scores $s(v^f)$ and select those whose scores exceed a predefined threshold $\lambda$, where $\lambda$ is a hyperparameter, which we tune with a step size of $0.05$ on the validation set and finally set to $0.45$. These selected nodes are considered to be strongly associated with the malicious prediction. For each selected function node, we extract its first-order subgraph to preserve the local semantic and structural context surrounding the corresponding function. The source code associated with the extracted subgraphs is then analyzed using LLMs, which enables fine-grained localization of malicious behaviors and provides human-interpretable explanations of the underlying attack activities.

\section{Evaluation}
\label{sec:evaluation}

In this section, we present a comprehensive evaluation of H$^2$GLM. We aim to answer the following research questions:

\begin{itemize}
    \item \textbf{RQ1 (Overall Effectiveness):} How does H$^2$GLM perform compared to state-of-the-art methods and large language models across packages of varying complexity?
    \item \textbf{RQ2 (Robustness):} How robust is H$^2$GLM when confronted with extreme cases, including minimally sized packages and those with extensive import dependencies?
    \item \textbf{RQ3 (Ablation Study):} What is the contribution of each key component in our proposed architecture?
    \item \textbf{RQ4 (Feature Representation):} Does the hierarchical heterogeneous graph structure effectively learn discriminative feature representations?
    \item \textbf{RQ5 (Malicious Behavior Localization):} Can H$^2$GLM identify and localize the specific functions responsible for malicious behavior?
\end{itemize}

\begin{table*}[t]
\centering
\caption{Comparative performance evaluation categorized by package file size. $N$ denotes the number of malicious samples in each category.}
\label{tab:package_size_results}
\renewcommand{\arraystretch}{1.15}
\setlength{\tabcolsep}{5.5pt}
\scriptsize
\begin{tabular}{l|cccc|cccc|cccc}
\hline
\multirow{2}{*}{\textbf{Method}} 
& \multicolumn{4}{c|}{\textbf{Small Packages} ($N=972$)} 
& \multicolumn{4}{c|}{\textbf{Large Packages} ($N=1162$)} 
& \multicolumn{4}{c}{\textbf{All Packages}} \\
\cline{2-13}
& \textbf{Acc} & \textbf{Prec} & \textbf{Rec} & \textbf{F1} 
& \textbf{Acc} & \textbf{Prec} & \textbf{Rec} & \textbf{F1} 
& \textbf{Acc} & \textbf{Prec} & \textbf{Rec} & \textbf{F1} \\
\hline
Clustering & 36.96 & 23.11 & 11.21 & 15.09 & 36.38 & 20.96 & 09.83 & 13.38 & 36.88 & 22.51 & 10.74 & 14.54 \\
Naive Bayes & 55.14 & 54.59 & 61.11 & 57.66 & 51.98 & 52.02 & 51.03 & 51.52 & 53.42 & 53.21 & 55.62 & 54.39 \\
Bandit4Mal & 48.15 & 42.10 & 26.50 & 32.53 & 41.20 & 35.80 & 18.20 & 24.13 & 44.37 & 39.06 & 21.98 & 28.13 \\
OSSGadget & 68.21 & 58.15 & 94.24 & 71.92 & 60.15 & 47.16 & 89.33 & 61.73 & 63.82 & 53.11 & \underline{91.57} & 67.23 \\
Warehouse & 79.78 & 94.61 & 63.17 & 75.76 & 63.94 & 89.32 & 31.67 & 46.76 & 71.16 & 92.55 & 46.02 & 61.47 \\
HERCULE & 92.54 & 93.30 & 91.67 & 92.48 & 86.14 & 83.28 & \underline{90.45} & \underline{86.72} & \underline{89.06} & 87.60 & 91.00 & \underline{89.27} \\
MPHunter & 91.10 & 86.20 & 93.50 & 89.70 & \underline{86.40} & 81.10 & 87.40 & 84.13 & 88.54 & 83.56 & 90.18 & 86.74 \\
\hline
Qwen2.5-7b & 48.17 & 47.93 & 42.55 & 45.08 & 42.51 & 40.29 & 31.07 & 35.09 & 45.40 & 44.39 & 36.42 & 40.01 \\
Llama3-8b & 45.30 & 46.79 & 68.58 & 55.63 & 43.08 & 45.01 & 62.43 & 52.31 & 43.72 & 45.61 & 65.20 & 53.67 \\
DeepSeek-R1-70b & 83.43 & 76.19 & \underline{97.25} & 85.44 & 57.25 & 60.08 & 43.22 & 50.27 & 70.13 & 71.24 & 67.52 & 69.33 \\
Qwen3-14b & \underline{94.84} & \textbf{98.03} & 91.51 & \underline{94.66} & 77.92 & \textbf{96.84} & 57.72 & 72.33 & 85.68 & \textbf{97.85} & 72.96 & 83.59 \\
\hline
\textbf{H$^2$GLM} & \textbf{97.31} & \underline{95.68} & \textbf{99.08} & \textbf{97.35} & \textbf{96.89} & \underline{95.51} & \textbf{98.40} & \textbf{96.93} & \textbf{97.08} & \underline{95.59} & \textbf{98.71} & \textbf{97.12} \\
\hline
\end{tabular}
\end{table*}

\begin{table*}[t]
\centering
\caption{Comparative performance evaluation categorized by import dependency complexity. $N$ indicates the number of malicious samples in each category.}
\label{tab:import_complexity_results}
\scriptsize
\renewcommand{\arraystretch}{1.25}
\setlength{\tabcolsep}{5.5pt}
\begin{tabular}{l|cccc|cccc|cccc}
\hline
\multirow{2}{*}{\textbf{Method}} 
& \multicolumn{4}{c|}{\textbf{Small Import Amounts} ($N=1647$)} 
& \multicolumn{4}{c|}{\textbf{Large Import Amount} ($N=487$)} 
& \multicolumn{4}{c}{\textbf{All Packages}} \\
\cline{2-13}
& \textbf{Acc} & \textbf{Prec} & \textbf{Rec} & \textbf{F1} 
& \textbf{Acc} & \textbf{Prec} & \textbf{Rec} & \textbf{F1} 
& \textbf{Acc} & \textbf{Prec} & \textbf{Rec} & \textbf{F1} \\
\hline
Clustering & 45.12 & 42.54 & 27.80 & 33.63 & 34.65 & 15.05 & 06.61 & 09.19 & 36.88 & 22.51 & 10.74 & 14.54 \\
Naive Bayes & 55.80 & 56.10 & 63.50 & 59.57 & 45.38 & 41.48 & 29.00 & 34.14 & 53.42 & 53.21 & 55.62 & 54.39 \\
Bandit4Mal & 46.80 & 41.20 & 23.60 & 30.01 & 36.14 & 28.16 & 16.50 & 20.81 & 44.37 & 39.06 & 21.98 & 28.13 \\
OSSGadget & 66.45 & 56.32 & 93.80 & 70.38 & 54.92 & 44.20 & 84.00 & 57.92 & 63.82 & 53.11 & \underline{91.57} & 67.23 \\
Warehouse & 70.31 & 92.94 & 43.96 & 59.69 & 74.44 & 91.61 & 53.80 & 67.79 & 71.16 & 92.55 & 46.02 & 61.47 \\
HERCULE & 90.22 & 90.97 & 89.31 & \underline{90.13} & \underline{85.11} & 78.50 & \textbf{96.71} & 86.66 & \underline{89.06} & 87.60 & 91.00 & \underline{89.27} \\
MPHunter & \underline{91.15} & 86.10 & \underline{94.00} & 89.88 & 79.72 & 74.80 & 77.20 & \underline{75.98} & 88.54 & 83.56 & 90.18 & 86.74 \\
\hline
Qwen2.5-7b & 47.13 & 46.69 & 40.43 & 43.34 & 35.40 & 29.00 & 20.16 & 23.78 & 45.40 & 44.39 & 36.42 & 40.01 \\
Llama3-8b & 50.64 & 50.42 & 76.91 & 60.91 & 20.80 & 17.53 & 15.76 & 16.60 & 43.72 & 45.61 & 65.20 & 53.67 \\
DeepSeek-R1-70b & 73.85 & 72.86 & 76.02 & 74.41 & 47.42 & 46.38 & 33.07 & 38.61 & 70.13 & 71.24 & 67.52 & 69.33 \\
Qwen3-14b & 90.05 & \textbf{97.87} & 81.89 & 89.17 & 67.83 & \textbf{95.39} & 37.47 & 53.80 & 85.68 & \textbf{97.85} & 72.96 & 83.59 \\
\hline
\textbf{H$^2$GLM} & \textbf{97.29} & \underline{95.52} & \textbf{99.23} & \textbf{97.34} & \textbf{95.84} & \underline{95.14} & \underline{96.62} & \textbf{95.88} & \textbf{97.08} & \underline{95.59} & \textbf{98.71} & \textbf{97.12} \\
\hline
\end{tabular}
\end{table*}

\subsection{Experimental Setup}
\label{subsec:setup}

\subsubsection{Dataset Construction}

To ensure the integrity of our evaluation and strictly prevent data leakage, we constructed our dataset under a rigorous decontamination protocol. By leveraging cryptographic hashing with MD5 alongside fuzzy hashing techniques, we enforced strict sample uniqueness throughout the dataset.

\noindent \textbf{Malicious Samples.}
Such samples are derived from a recent large-scale benchmark for malicious Python package detection~\cite{gao2026efficientcodeanalysisgraphguided}, supplemented by packages we collected from public security advisories and threat intelligence feeds. After deduplication, we obtained a total of 3,049 unique malicious packages, which are available in the replication package linked in the Abstract.

\noindent \textbf{Benign Samples.}
The benign samples were collected from the most popular packages on the Python Package Index. We selected packages with high download counts and stratified the selection to match the file size distribution of the malicious samples.

\noindent \textbf{Data Splitting.}
We adopt a 15\%/15\%/70\% train/val/test split to simulate realistic scenarios with scarce labeled malicious samples, yielding 2,134 test packages that serve as the basis for all subsequent evaluations.

\subsubsection{Baselines and Comparison Methods}

We compared our method against two categories of baselines.

\noindent \textbf{Static Analysis and Machine Learning Methods.}
\begin{itemize}
    \item \textbf{Clustering:} A K-Means clustering approach trained on TF-IDF features extracted from the source code.
    \item \textbf{Naive Bayes:} A probabilistic classifier using code statistical features and bag-of-words representations.
    \item \textbf{Bandit4Mal}~\cite{bandit4mal}: A security-oriented variant of the Bandit static analysis tool customized for detecting malicious patterns.
    \item \textbf{OSSGadget}~\cite{ossgadget}: A widely adopted signature-based scanner for open-source supply chain security.
    \item \textbf{Warehouse}~\cite{warehouse}: PyPI's official malware
    detection system that applies YARA rules to \texttt{setup.py} files,
    flagging known dangerous constructs such as process-spawning calls
    and network-access patterns.
    \item \textbf{HERCULE}~\cite{shariffdeen2025hercule}: A
    dynamic analysis framework that executes packages in isolated
    containers and detects malicious behaviors through system call
    monitoring during installation.
    \item \textbf{MPHunter}~\cite{liang2023mphunter}: A state-of-the-art graph-based malware detection method that utilizes message passing on function call graphs.
\end{itemize}

\noindent \textbf{Large Language Models.}
We evaluated four leading LLMs to assess their zero-shot malware detection capabilities: \textbf{Qwen2.5-7b}~\cite{qwen2.5}, \textbf{Llama3-8b}~\cite{grattafiori2024llama}, \textbf{Qwen3-14b}~\cite{qwen3technicalreport}, and the reasoning-enhanced \textbf{DeepSeek-R1-70b}~\cite{deepseekai2025deepseekr1incentivizingreasoningcapability}.

Since LLMs cannot directly process compressed archives, we pre-processed the packages by recursively traversing the directory tree and extracting all Python scripts. File paths are inserted as comments before each file's content to preserve structural context. The combined text is truncated if it exceeds the model's maximum context window, prioritizing \texttt{setup.py} and \texttt{\_\_init\_\_.py} files as they are common entry points for malicious payloads.

We designed a structured prompt that instructs the model to act as a security analyst, performing static analysis, heuristic evaluation of risky functions such as \texttt{eval}, \texttt{exec}, and \texttt{subprocess}, and intent verification to distinguish between legitimate and malicious usage patterns. The prompt enforces a specific output format requiring the model to provide detailed reasoning before the final binary prediction, ensuring consistent and parseable responses across all evaluated models.

\subsubsection{Evaluation Metrics}

We adopt four standard metrics for binary classification: \textbf{Accuracy}, \textbf{Precision}, \textbf{Recall}, and \textbf{F1-Score}. All metrics are reported as percentages. The best results are highlighted in \textbf{bold}, and the second-best results are \underline{underlined}.

To comprehensively evaluate detection performance, we stratified the test set along two orthogonal dimensions: package file size and import dependency complexity. This stratification reveals how different methods handle packages of varying structural complexity.

\begin{table*}[t]
\centering
\caption{Robustness evaluation on extreme package size categories. $N$ denotes the number of malicious samples.}
\label{tab:extreme_size_results}
\scriptsize
\renewcommand{\arraystretch}{1.15}
\setlength{\tabcolsep}{5.5pt}
\begin{tabular}{l|cccc|cccc}
\hline
\multirow{2}{*}{\textbf{Method}} 
& \multicolumn{4}{c|}{\textbf{Extremely Small} ($N=136$, $\leq$1.57KB)} 
& \multicolumn{4}{c}{\textbf{Extremely Large} ($N=109$, $\geq$700KB)} \\
\cline{2-9}
& \textbf{Acc} & \textbf{Prec} & \textbf{Rec} & \textbf{F1} 
& \textbf{Acc} & \textbf{Prec} & \textbf{Rec} & \textbf{F1} \\
\hline
Clustering & 45.21 & 43.40 & 31.51 & 36.51 & 41.53 & 35.29 & 20.34 & 25.81 \\
Naive Bayes & 50.46 & 51.02 & 45.45 & 48.08 & 47.70 & 46.88 & 27.27 & 34.48 \\
Bandit4Mal & 52.21 & 52.24 & 51.47 & 51.85 & 49.54 & 50.00 & 54.55 & 52.17 \\
OSSGadget & 69.85 & 65.17 & 85.29 & 73.89 & 65.14 & 69.77 & 54.55 & 61.22 \\
Warehouse & 54.41 & 87.50 & 10.29 & 18.42 & 85.32 & \underline{96.39} & 73.39 & 83.33 \\
HERCULE & 87.50 & 83.12 & 94.12 & 88.28 & 78.44 & 72.79 & 90.83 & 80.82 \\
MPHunter & \underline{92.65} & \underline{90.28} & 95.59 & \underline{92.86} & \underline{89.91} & 89.29 & \underline{90.91} & \underline{90.09} \\
\hline
Qwen2.5-7b & 44.85 & 44.53 & 41.91 & 43.18 & 41.82 & 40.82 & 36.36 & 38.46 \\
Llama3-8b & 41.36 & 42.75 & 50.91 & 46.47 & 35.66 & 36.91 & 40.44 & 38.60 \\
DeepSeek-R1-70b & 80.51 & 72.68 & \textbf{97.79} & 83.39 & 40.45 & 32.20 & 17.27 & 22.49 \\
Qwen3-14b & 90.77 & 89.55 & 92.31 & 90.91 & 67.27 & \textbf{97.50} & 35.45 & 52.00 \\
\hline
\textbf{H$^2$GLM} & \textbf{95.96} & \textbf{95.62} & \underline{96.32} & \textbf{95.97} & \textbf{96.33} & 95.50 & \textbf{97.25} & \textbf{96.36} \\
\hline
\end{tabular}
\end{table*}

\begin{table*}[t]
\centering
\caption{Robustness evaluation on extreme import dependency categories. $N$ denotes the number of malicious samples.}
\label{tab:extreme_import_results}
\scriptsize
\renewcommand{\arraystretch}{1.15}
\setlength{\tabcolsep}{5.5pt}

\begin{tabular}{l|cccc|cccc}
\hline
\multirow{2}{*}{\textbf{Method}} 
& \multicolumn{4}{c|}{\textbf{Minimal Imports} ($N=103$, 0--1 imports)} 
& \multicolumn{4}{c}{\textbf{Extensive Imports} ($N=90$, $>$100 imports)} \\
\cline{2-9}
& \textbf{Acc} & \textbf{Prec} & \textbf{Rec} & \textbf{F1} 
& \textbf{Acc} & \textbf{Prec} & \textbf{Rec} & \textbf{F1} \\
\hline
Clustering & 39.53 & 23.53 & 09.30 & 13.33 & 46.81 & 44.44 & 25.53 & 32.43 \\
Naive Bayes & 52.13 & 62.50 & 10.64 & 18.18 & 29.79 & 22.86 & 17.02 & 19.51 \\
Bandit4Mal & 45.15 & 44.44 & 38.83 & 41.45 & 30.56 & 29.41 & 27.78 & 28.57 \\
OSSGadget & 69.90 & 67.23 & 77.67 & 72.07 & 29.44 & 17.78 & 08.89 & 11.85 \\
Warehouse & 59.71 & \textbf{100.00} & 19.42 & 32.52 & 67.22 & \underline{91.89} & 37.78 & 53.54 \\
HERCULE & 63.11 & 68.00 & 49.51 & 57.30 & \underline{82.22} & 75.89 & \textbf{94.44} & \underline{84.16} \\
MPHunter & \underline{92.23} & 93.94 & \underline{90.29} & \underline{92.08} & 40.56 & 41.24 & 44.44 & 42.78 \\
\hline
Qwen2.5-7b & 43.69 & 43.30 & 40.78 & 42.00 & 29.67 & 19.67 & 13.19 & 15.79 \\
Llama3-8b & 49.51 & 49.64 & 66.99 & 57.02 & 34.07 & 34.41 & 35.16 & 34.78 \\
DeepSeek-R1-70b & 65.05 & 64.76 & 66.02 & 65.38 & 31.87 & 18.87 & 10.99 & 13.89 \\
Qwen3-14b & 88.83 & \underline{98.78} & 78.64 & 87.57 & 49.45 & 49.18 & 32.97 & 39.46 \\
\hline
\textbf{H$^2$GLM} & \textbf{96.60} & 95.28 & \textbf{98.06} & \textbf{96.65} & \textbf{95.00} & \textbf{95.51} & \textbf{94.44} & \textbf{94.97} \\
\hline
\end{tabular}
\end{table*}

\subsection{RQ1: Overall Effectiveness}
\label{subsec:rq1}
\noindent \textbf{Stratification by Package Size.}
We categorized packages based on the cumulative size of Python scripts within the archive. Packages with a total script size below 5KB are classified as \textit{Small Packages}, while the remainder are classified as \textit{Large Packages}. Table~\ref{tab:package_size_results} presents the comparative results.

\noindent \textbf{Stratification by Import Complexity.}
We further stratified packages based on the number of import statements. Packages containing fewer than 10 import statements are classified as \textit{Small Import Amount}, while those with 10 or more are classified as \textit{Large Import Amount}. Table~\ref{tab:import_complexity_results} presents the results.

\noindent \textbf{Analysis.}
The results demonstrate that H$^2$GLM consistently outperforms all baseline methods across both stratification dimensions as well as on the combined dataset. Signature-based methods such as OSSGadget achieve high recall but suffer from low precision, indicating a tendency toward false positives. This behavior stems from their reliance on predefined patterns that may match benign code constructs. On the other hand, LLMs exhibit inconsistent performance across package sizes. Notably, Qwen3-14b achieves 94.66\% F1 on small packages but drops to 72.33\% on Large Packages. This degradation arises from context window limitations that force truncation of larger codebases, causing the model to miss critical malicious segments. H$^2$GLM maintains stable performance across all categories with less than 2\% variance in F1-Score, demonstrating the effectiveness of our heterogeneous hierarchical graph representation in handling diverse package structures.

\subsection{RQ2: Robustness on Extreme Cases}
\label{subsec:rq2}

To stress-test the robustness of detection methods, we constructed extreme subsets representing boundary conditions in real-world scenarios.

\noindent \textbf{Extreme Size Categories.}
Packages with cumulative Python script size at most 1.57KB are classified as \textit{Extremely Small}, representing minimal attack payloads. Packages with script size at least 700KB are classified as \textit{Extremely Large}, representing sophisticated attacks embedded within substantial codebases. Table~\ref{tab:extreme_size_results} presents the results.

\noindent \textbf{Extreme Import Categories.}
Packages containing zero or one import statement are classified as \textit{Minimal Imports}, representing self-contained attacks. Packages with more than 100 import statements are classified as \textit{Extensive Imports}, representing attacks that leverage complex dependency chains. Table~\ref{tab:extreme_import_results} presents the results.

\noindent \textbf{Analysis.}
The extreme-case evaluation further highlights the limitations of existing methods. First, all baselines exhibit noticeable performance degradation on the \textit{Extensive Imports} category. For example, MPHunter drops from 92.08\% F1 on \textit{Minimal Imports} to 42.78\% F1 on \textit{Extensive Imports}, indicating that homogeneous graph representations struggle to capture complex dependency structures. Second, LLMs perform well on small packages but degrade substantially on large codebases. For instance, DeepSeek-R1-70b achieves 83.39\% F1 on extremely small packages but only 22.49\% on extremely large ones, suggesting that long-context reasoning remains challenging. In contrast, H$^2$GLM consistently achieves F1-scores between 94.97\% and 96.65\% across all extreme settings, demonstrating the robustness of the proposed hierarchical heterogeneous graph representation.

\subsection{RQ3: Ablation Study}
\label{subsec:rq3}

\begin{figure*}[h]
    \centering
    \begin{subfigure}[b]{0.24\textwidth}
        \centering
        \includegraphics[width=\textwidth]{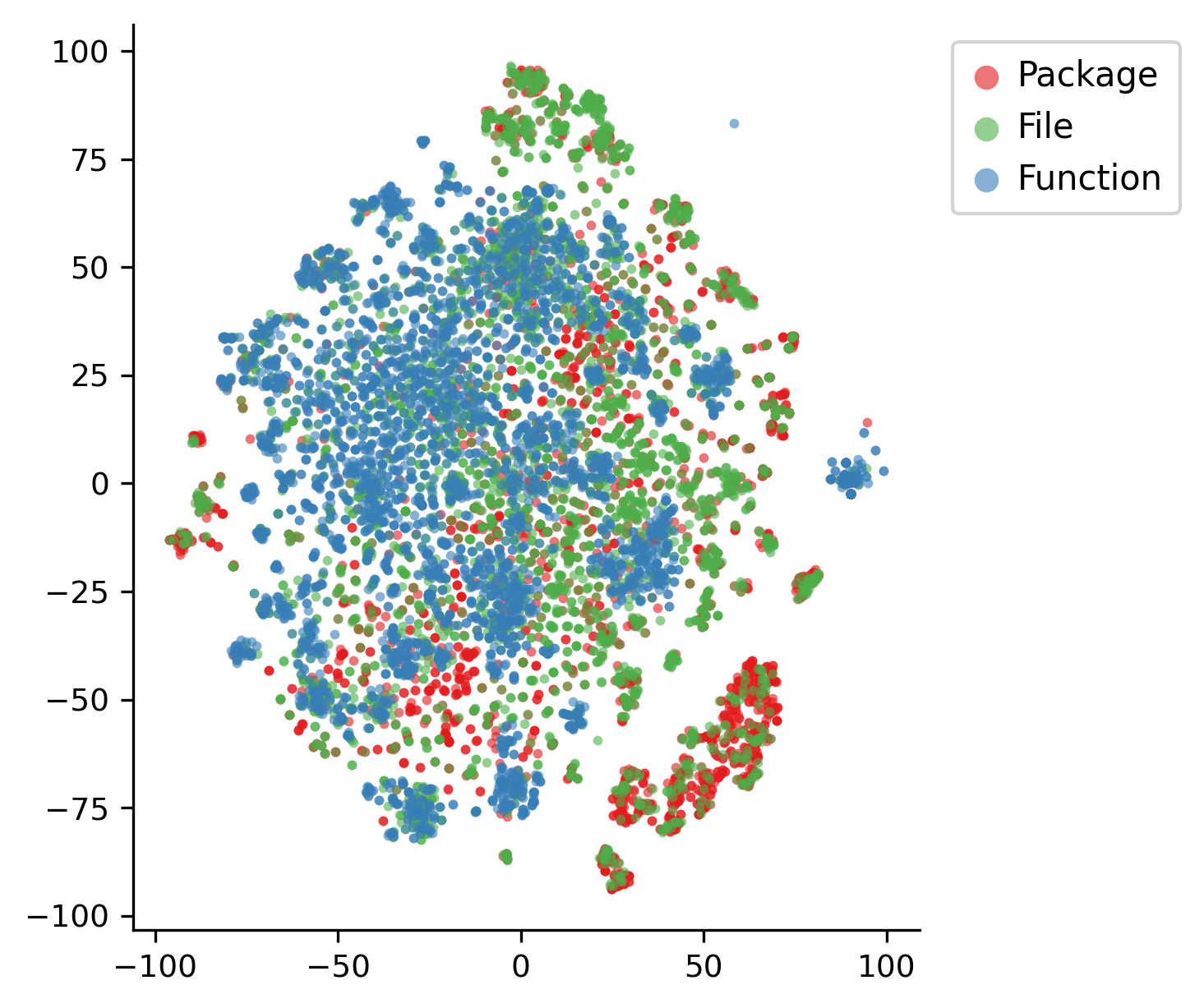}
        \caption{Raw features}
        \label{fig:tsne_raw}
    \end{subfigure}
    \hfill
    \begin{subfigure}[b]{0.24\textwidth}
        \centering
        \includegraphics[width=\textwidth]{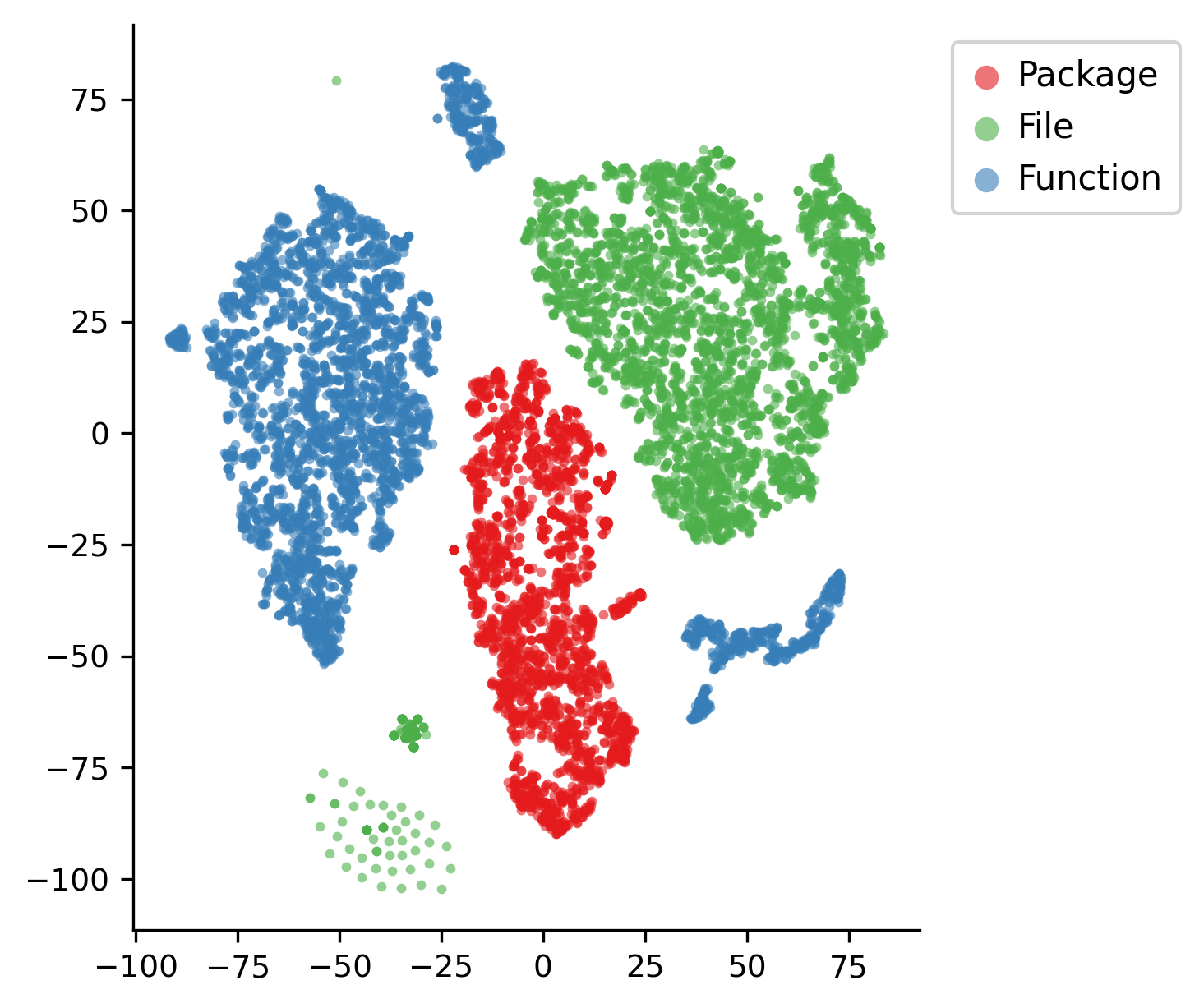}
        \caption{After type alignment}
        \label{fig:tsne_aligned}
    \end{subfigure}
    \hfill
    \begin{subfigure}[b]{0.24\textwidth}
        \centering
        \includegraphics[width=\textwidth]{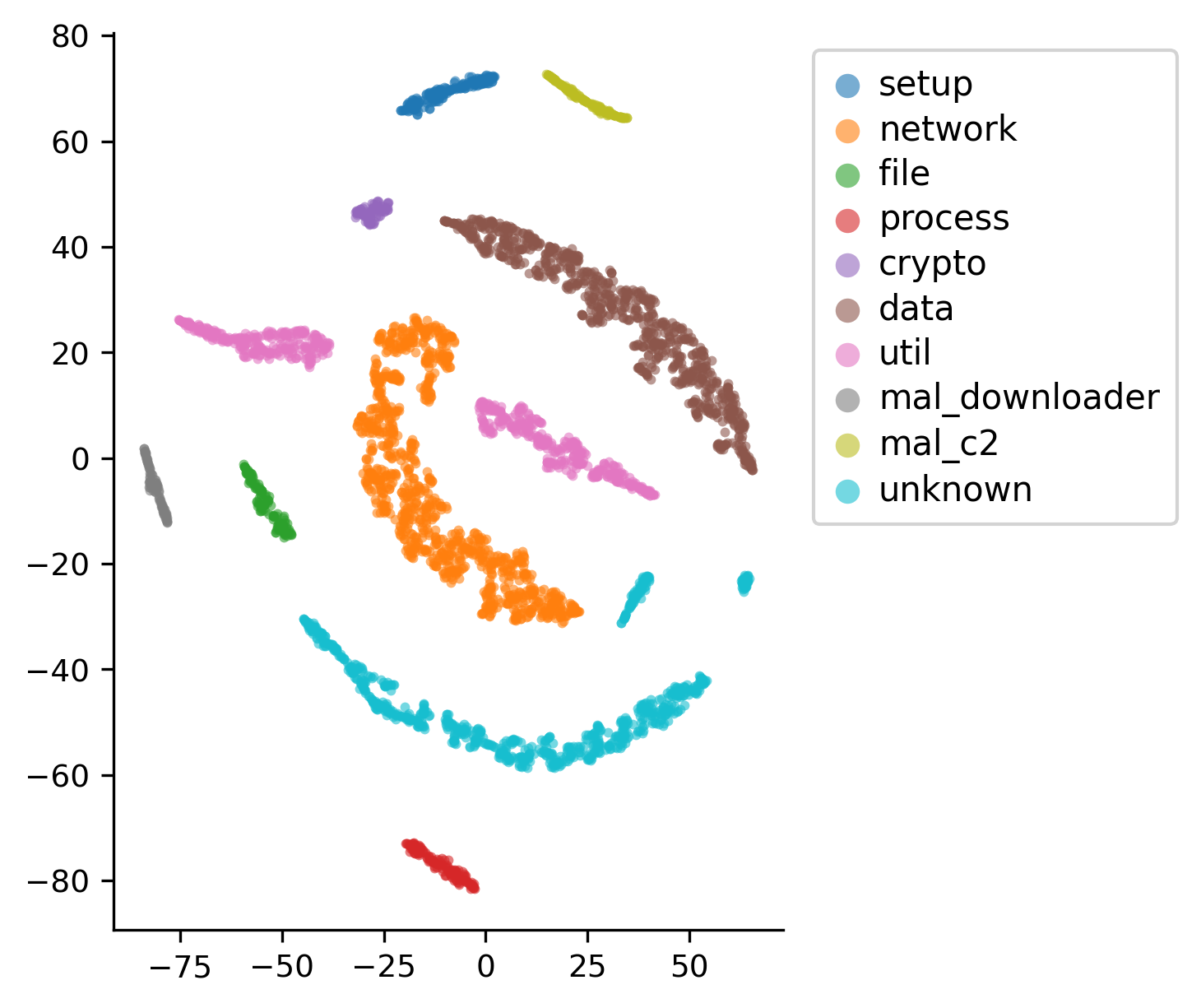}
        \caption{After content handling}
        \label{fig:tsne_content}
    \end{subfigure}
    \hfill
    \begin{subfigure}[b]{0.24\textwidth}
        \centering
        \includegraphics[width=\textwidth]{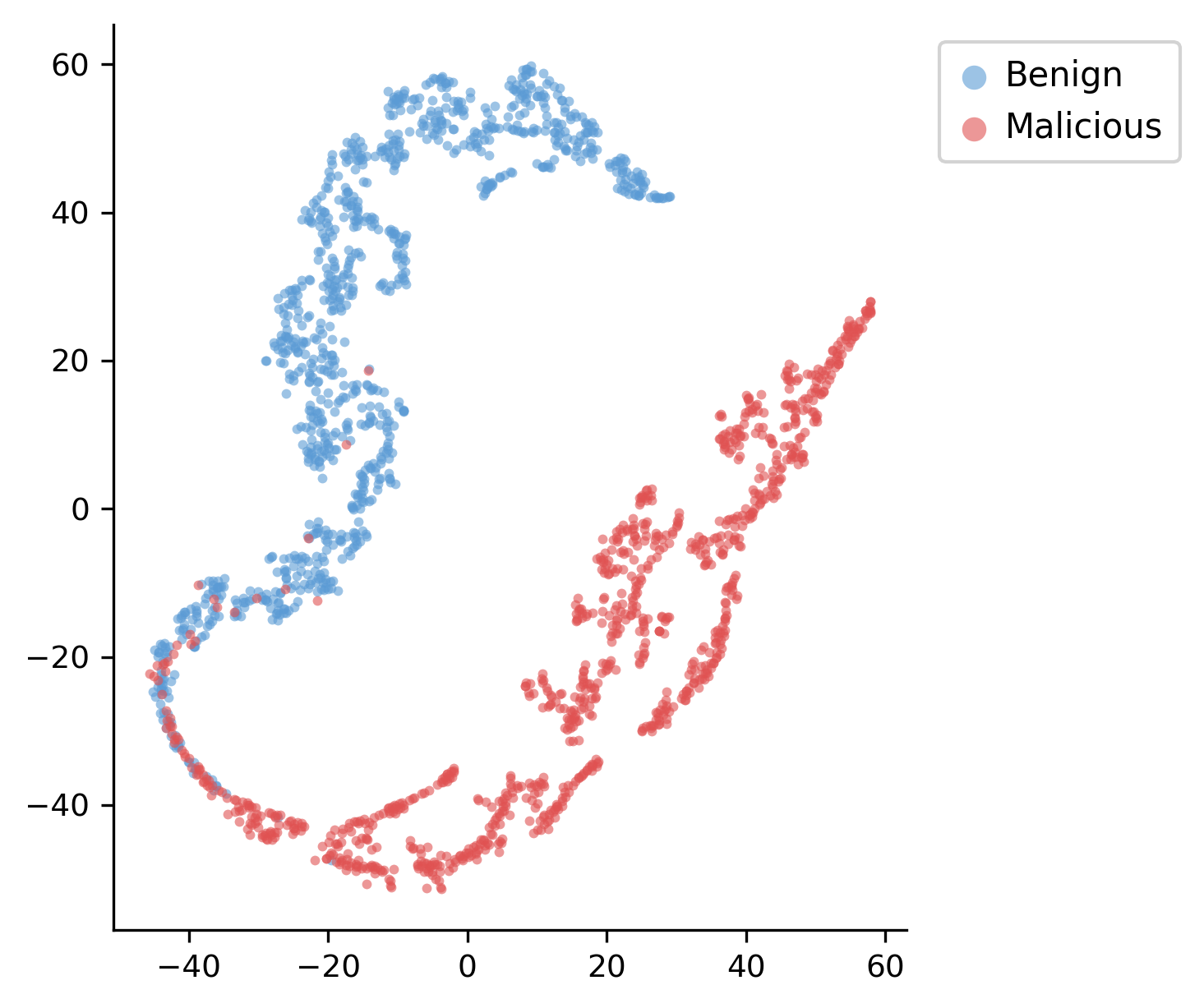}
        \caption{Final embeddings}
        \label{fig:tsne_final}
    \end{subfigure}
    \caption{T-SNE visualization of feature representations at different stages.}
    \label{fig:tsne}
    \vskip -0.1in
\end{figure*}

\begin{table}[b]
\centering
\caption{Ablation study results, where the full model achieves optimal 
performance. ``w/o All Components'' retains only the base GNN 
(no $\phi^{\text{LLM}}(v)$-guided processing, $\psi(e)$-guided message 
passing, or CodeBERT), and ``LLM Feature Extractor'' replaces CodeBERT 
with LLM-generated embeddings.}
\label{tab:ablation}

\renewcommand{\arraystretch}{1.15}
\setlength{\tabcolsep}{5.5pt}
\scriptsize
\begin{tabular}{l|cccc}
\hline
\textbf{Model Variant} 
& \textbf{Acc} 
& \textbf{Prec} 
& \textbf{Rec} 
& \textbf{F1} \\
\hline
\textbf{H$^2$GLM (Full Model)} & \textbf{97.08} & \textbf{95.59} & 98.71 & \textbf{97.12} \\
\hline
w/o $\psi(e)$-Guided Message Passing & \underline{92.76} & 88.37 & 98.47 & \underline{93.15} \\
w/o $\phi^{\text{LLM}}(v)$-Guided Processing & 91.75 & 86.37 & \textbf{99.16} & 92.32 \\
w/o CodeBERT & 92.60 & 87.87 & 98.84 & 93.03 \\
\hline
w/o $\phi^{\text{LLM}}(v)$ \& $\psi(e)$ Components & 89.95 & \underline{89.15} & 91.20 & 90.16 \\
w/o All Components & 86.70 & 86.10 & 87.50 & 86.79 \\
\hline
LLM Feature Extractor & 92.62 & 87.80 & \underline{99.00} & 93.07 \\
\hline
\end{tabular}
\end{table}

To systematically evaluate the contributions of each component in H$^2$GLM, we conducted ablation studies by progressively removing key modules. Table~\ref{tab:ablation} presents the results.

\noindent \textbf{Analysis.}
The ablation results reveal that removing the $\psi(e)$-guided message passing component causes a 3.97\% drop in F1-Score. This degradation confirms that heterogeneous message passing, which distinguishes between node types and edge types, is essential for capturing the hierarchical semantics of package structures. Furthermore, removing the $\phi^{\text{LLM}}(v)$-guided processing component results in a 4.80\% F1 decrease. This module classifies function nodes into semantic categories such as network operations, file manipulations, and cryptographic routines, providing critical behavioral context for maliciousness determination. The LLM Feature Extractor variant, which replaces CodeBERT with LLM-generated embeddings, achieves comparable but slightly inferior performance. This suggests that task-specific pretrained models are more effective for feature extraction in our model.

\subsection{RQ4: Feature Representation Analysis}
\label{subsec:rq4}

To validate that our hierarchical heterogeneous graph structure learns discriminative representations, we visualize the feature evolution across different stages using t-SNE projections. Figure~\ref{fig:tsne} presents the visualization results.

The t-SNE visualization confirms a clear progression: nodes separate by type after alignment, semantic clusters emerge after content handling, and final embeddings show clean malicious/benign separation.

\subsection{RQ5: Malicious Behavior Localization}
\label{subsec:rq5}
We evaluate the function-level localization capability described in Section~\ref{subsec:Localization} through quantitative experiments and case study analysis.

\subsubsection{Quantitative Evaluation}

Since no publicly available benchmark provides function-level annotations for malicious Python packages, we manually established the localization ground truth for a randomly sampled subset of 300 malicious packages. Four authors independently inspected the source code and identified the function(s) responsible for the primary malicious behavior according to a predefined annotation protocol. Whenever publicly available malware analyses or security advisories were available, they were consulted as supporting evidence during the annotation process. Disagreements were resolved through discussion until consensus was reached. Our method successfully identified the actual malicious functions in 287 out of 300 packages, achieving a localization accuracy of 95.67\%. 

\subsubsection{Case Study}

We present a detailed case study on \texttt{cipherbcrypt-1.4}, a malicious package masquerading as a legitimate cryptographic library. The package contains 3 functions with an original malicious probability of 0.7295. Table~\ref{tab:dropnode_case} presents the perturbation analysis results.

\begin{table}[htbp]
\scriptsize
\centering
\caption{DropNode analysis for \texttt{cipherbcrypt-1.4}. $\Delta$Prob denotes the probability drop after removing each function.}
\label{tab:dropnode_case}
\scriptsize
\begin{tabular}{lccc}
\toprule
\textbf{Function Name} & \textbf{Type} & \textbf{Prob} & \textbf{$\Delta$Prob} \\
\midrule
\texttt{algorithmb.ciphersd} & network & 0.4514 & \textbf{0.2781} \\
\texttt{algorithmb.decd} & crypto & 0.7196 & 0.0099 \\
\texttt{algorithmb.encd} & crypto & 0.7198 & 0.0097 \\
\bottomrule
\end{tabular}
\end{table}

Removing \texttt{algorithmb.ciphersd} causes the malicious probability to drop from 0.7295 to 0.4514, changing the prediction from malicious to benign. The other two functions have negligible impact, precisely identifying \texttt{ciphersd} as the core malicious function.

The identified function implements a command-and-control communication channel with heavy obfuscation:

\begin{lstlisting}[language=Python, basicstyle=\ttfamily\scriptsize, breaklines=true,
  keywordstyle=\color{blue!80!black}\bfseries,
  commentstyle=\color{green!50!black}\itshape,
  stringstyle=\color{red!70!black},
  emphstyle=\color{purple}]
def ciphersd(self, e:str, t:int, v:str):
  try:
    gG53z=[b'=QMBn+BAF8szLNNzK3USuwCSrwJe',
           b'kQQqdAAA23wcw0iCLe31so0LTzJe']
    doSGb=HTTPSConnection(self.decd(gG53z[0]),timeout=1)
    doSGb.request(''.join(map(chr,[71,69,84])),
                  self.decd(gG53z[1]))
    tenNo=doSGb.getresponse().read().decode().strip()
    Fd3hh=self.decd(tenNo).split(':')
    if int(Fd3hh[3])==0:
      pF3th={'t':t,'n':getlogin(),'v':v,'e':self.encd(e)}
      s0Ntz=HTTPSConnection(Fd3hh[0],Fd3hh[1],timeout=1)
      s0Ntz.request(''.join(map(chr,[80,79,83,84])),
                    Fd3hh[2],dumps(pF3th),headers)
      return True
  except:
    return False
\end{lstlisting}

The function employs multiple obfuscation techniques: variable names are randomized strings, HTTP methods are constructed via \texttt{chr()} sequences where \texttt{[71,69,84]} decodes to ``GET'' and \texttt{[80,79,83,84]} decodes to ``POST'', and server addresses are stored as reversed base64-encoded strings decoded at runtime. The attack first retrieves C2 server configuration from a hardcoded initial server, then exfiltrates sensitive data including the username via \texttt{getlogin()}. The two supporting functions \texttt{encd} and \texttt{decd} provide obfuscated encoding utilities that appear benign in isolation but enable the data exfiltration pipeline.

This case demonstrates that despite heavy obfuscation, our method correctly identifies the network communication function as the primary source of maliciousness, with the probability drop of 0.2781 providing a clear quantitative signal distinguishing it from supporting utilities.

\subsection{Comparison with Homogeneous GNNs}
\label{subsec:gnn_comparison}

To validate the necessity of heterogeneous graph modeling, we compare H$^2$GLM against standard homogeneous GNN architectures including Graph Convolutional Network and Graph Attention Network. Both baselines operate exclusively on function-level features extracted by CodeBERT, with function call relationships serving as edges. This setup foregoes the hierarchical package-file-function structure in favor of a simpler function-centric graph. Both models adopt a two-layer architecture followed by global mean pooling and an MLP classifier, with hidden dimension set to 128 and dropout probability of 0.3.

\begin{table}[h]
\scriptsize
\centering
\caption{Comparison with homogeneous GNNs.}
\label{tab:gnn_comparison}
\begin{tabular}{lcccc}
\toprule
\textbf{Method} & \textbf{Acc} & \textbf{Prec} & \textbf{Rec} & \textbf{F1} \\
\midrule
GCN & 81.45 & 91.18 & 70.64 & 79.61 \\
GAT & 85.17 & 80.81 & 93.29 & 86.60 \\
\midrule
\textbf{H$^2$GLM} & \textbf{97.08} & \textbf{95.59} & \textbf{98.71} & \textbf{97.12} \\
\bottomrule
\end{tabular}
\end{table}

Table~\ref{tab:gnn_comparison} presents the results. Both homogeneous GNN baselines significantly underperform compared to H$^2$GLM, with F1-Score gaps of 17.51\% and 10.52\% for GCN and GAT respectively. GCN exhibits high precision but low recall due to its uniform aggregation mechanism that dilutes malicious signals when malicious functions constitute a small fraction of total functions. GAT achieves better recall by leveraging attention mechanisms, but without explicit type information, it cannot effectively distinguish between different roles of functions. These results confirm that the hierarchical heterogeneous structure is essential for effective malware detection.

\subsection{Real-World Deployment}
\label{subsec:deployment}
To evaluate the practical effectiveness of H$^2$GLM in real-world scenarios, we deployed our system to continuously monitor newly uploaded packages on PyPI. Over a four-month observation period, our system scanned over 20,000 newly published packages. We submitted 32 flagged packages as potential malware reports to the PyPI security team, of which 19 were confirmed and subsequently removed by the maintainers, yielding an acceptance rate of 59.4\%.The 13 unconfirmed cases stem from outbound network communication shared by both malware and legitimate SDKs; incorporating service registration context could help mitigate such false positives. 

Notably, the entire detection pipeline operates in a fully automated manner—from raw package ingestion and heterogeneous graph construction to LLM-enhanced semantic extraction and final classification—requiring zero human intervention, demonstrating H$^2$GLM's practical suitability for continuous, large-scale PyPI ecosystem monitoring.

\section{Threats to Validity} 

\paragraph{Generalizability of the experimental results} the effectiveness of H$^2$GLM depends on the representativeness of the evaluation dataset. To mitigate this threat, we evaluate the proposed framework on real-world malicious and benign PyPI packages, comparing it with diverse baselines under multiple settings, including different package sizes, import complexities, and localization tasks. Although our evaluation focuses on the PyPI ecosystem, the proposed framework can be readily extended to other software ecosystems. 

\paragraph{Implementation of the proposed framework} the performance of learning-based methods may vary with implementation details and hyperparameter settings. To reduce this threat, all methods are evaluated under the same experimental protocol, and model parameters are selected based on validation performance. The consistent results across different evaluation settings indicate the robustness of the proposed framework.

\section{Related Works}
\label{sec:related_work}

\textit{Software supply chain security:} Software supply chain attacks have been widely studied, with common threats including typosquatting, dependency confusion, and malicious installation scripts~\cite{vu2020typosquatting,biondi2021dependency,ohm2020backstabber}. Surveys show that deep learning has become the dominant detection approach, but remains vulnerable to obfuscation and zero-day attacks~\cite{berrios2025systematic,alshoulie2025deep}.

\textit{Static, dynamic, and kernel-level analysis:} Static analysis methods such as GuardDog and Bandit4Mal are efficient but rely on handcrafted rules. Dynamic analysis (e.g., MalOSS) improves behavioral visibility via sandbox execution but suffers from high overhead and incomplete coverage. Kernel-level methods based on eBPF provide lightweight monitoring but remain limited by platform dependence and evasion strategies such as anti-analysis and delayed execution~\cite{duan2021towards,bogle2023ebpf,pierazzi2020intriguing}.

\textit{GNNs for code analysis:} GNN-based methods model code using AST, CFG, and DFG representations. Representative works such as Devign and MPHunter demonstrate strong performance in vulnerability and malware detection~\cite{zhou2019devign,liang2023mphunter}. Recent heterogeneous GNNs further improve representation capacity by modeling different node and edge types~\cite{wang2019heterogeneous,hu2020heterogeneous,fu2020magnn}, but still struggle with complex malicious behavior propagation. Recent research has begun to integrate GNNs and LLMs for code security analysis~\cite{li2026pvdetector,yamaguchi2014cpg}.

\textit{LLMs for code security:} Pretrained code LLMs such as CodeBERT and CodeLlama provide strong semantic understanding for security tasks~\cite{feng2020codebert,roziere2023code}. However, they suffer from distribution shift and scalability limitations in real-world settings~\cite{ding2024vulnerability}. Recent approaches combine LLMs or LMs with graph learning models, showing that integrating semantic reasoning with graph-based representations is a promising direction~\cite{gao2025malguard,linredetect}.

\section{Conclusions}

In this work, we propose H$^2$GLM, a hierarchical heterogeneous graph learning framework enhanced with LLMs for malicious Python package detection. By modeling the hierarchical heterogeneous information within the code graphs, the model captures both structural and semantic information for malware analysis. Experiments show that our method consistently outperforms strong baselines and demonstrates strong robustness across different package complexities.







\bibliographystyle{IEEEtran}
\bibliography{ref}

\end{document}